# Superconducting and critical properties of PrOFe$_{0.9}$Co$_{0.1}$As: effect of P doping

Shilpam Sharma[1], Jai Prakash[2], Gohil S. Thakur[2], A. T. Satya[1], A. Bharathi[*1],
A K Ganguli[*2] and C S Sundar[1]

[1]Material Science Group, Indira Gandhi Center for Atomic Research, Kalpakkam 603102 India

[2] Department of Chemistry, Indian Institute of Technology, New Delhi 110016 India

[*]corresponding authors:  bharathi@igcar.gov.in; ashok@chemistry.iitd.ac.in

**Abstract:** PrOFe$_{0.9}$Co$_{0.1}$As is known to be a superconductor with a T$_C$ of ~14 K. In an attempt to induce charge transfer and also induce chemical pressure between the FeAs and PrO layers we have synthesized for the first time oxypnictides of the type PrOFe$_{0.9}$Co$_{0.1}$As$_{1-x}$P$_x$ (P doping at As sites). All the compounds crystallize in the tetragonal ZrCuSiAs type structure (space group = P4/nmm). The lattice parameters (a and c) decrease as expected on substitution of smaller phosphorous at the arsenic site in PrOFe$_{0.9}$Co$_{0.1}$As and a decrease in T$_C$ was observed on substitution of P ions at a low rate of 0.13K/at. % of P for x > 0.1. The irreversibility field (H$_{irr}$) and critical current density (J$_C$), obtained using the AC susceptibility measurements was found to decrease monotonically with increasing 'P' concentration. 'H$_{irr}$' is observed to be much smaller than H$_{C2}$, pointing to a very low pinning energy. The pinning potential obtained using both in-field transport and AC susceptibility measurements indicate a low value of ~40 meV and show no significant variation with P substitution.

**Keywords:** Superconductivity, Critical Current Density (Jc), Irreversible field.

PACS: **74.70.Xa, 74.10.+v,74.25.Sv**

**I. INTRODUCTION**

After the discovery of superconductivity (SC) in quaternary oxypnictide compounds (LnO/FFeAs, Ln=rare earth) [1], much effort has been made to push the $T_C$ of these compounds beyond the 77 K benchmark. Several new compounds in all the four families (namely 1111, 122, 11 and 111) were discovered but the highest $T_C$ reported has remained at ~55 K for the Sm based 1111 compound [2]. Theoretical studies suggest that multiband superconductivity operates in these compounds arising primarily from the FeAs electronic bands [3, 4, 5, 6]. The normal state of these superconductors is well described by band theory. Fermi surface nesting leads to an itinerant anti-ferromagnetic ground state. Moving away from nesting by chemical substitution or applying mechanical pressure gives rise to spin fluctuations that favours the emergence of a superconducting ground state [3].

Although until now much emphasis was restricted to the basic understanding of these superconductors, recently there are reports of successful wire fabrication of these materials [7, 8]. Due to multiband effects, the upper critical field ($H_{C2}$) is reported to be higher for these compounds [9, 10, 11, 12]. These iron based superconductors show a low value of in-field intragranular pinning and weak link behaviour across grain boundaries [13,14] both of which limit their practical use. For practical application of these superconductors, it is important to understand how the flux dynamics and flux pinning determine the critical current density ($J_C$).

AC susceptibility measurements have been widely used in studies of HTSC particularly to determine, the critical current density, $J_C$, flux dynamics, irreversibility field ($H_{irr}$) and flux pinning potential [15, 16, 17]. The shift of the peak temperature in the out-of-phase component of temperature dependent complex susceptibility ($\chi''$), measured at different DC fields, has been widely used as a signature of the irreversibility line $H_{irr}(T)$ in cuprates [18, 19]. $H_{irr}(T)$ is a very important parameter to determine the technological worth of a SC material. The shift in the position of the peak in $\chi''(H)$ under different AC fields can be used to evaluate the $J_C$ of the compound, the particular DC field being close to the full penetration field. Similarly the peak shifts in $\chi''(T)$ observed with increasing frequency of AC field is also widely used to estimate the flux pinning energy in superconductors[20].

P substitution at As site, is known to induce superconductivity in $BaFe_2As_2$[21]. Here we report studies on P doping at As site in $PrOFe_{0.9}Co_{0.1}As$ which has a $T_C$ of ~14.2 K[22]. Since P is smaller than As, a lattice contraction is expected and the substitution should mimic the effect of the application of external pressure[21]. It was shown by very early measurements[23] that the $dT_C/dP$ is a strong function of the level of doping, viz., under-doped specimens showed $dT_C/dP >0$, while optimally doped samples show no change in $T_C$ under pressure and over doped samples show a decrease in $T_C$ with pressure. The present investigation is on P substitution effects on specimens that are optimally doped with Co, $T_C$ decreases with substitution, contrary to the expectation from the application of external pressure [23]. Further, based on the results of different AC susceptibility and resistivity measurements performed on P substituted samples, $H_{C2}$, $H_{irr}$, $J_C$ and pinning potential are obtained. Our findings indicate that P substitution leads to the degradation of the superconducting properties of the compound.

## II. Experimental Details

Polycrystalline samples with nominal compositions of $PrOFe_{0.9}Co_{0.1}As_{1-x}P_x$ (x = 0, 0.05, 0.1, 0.15, 0.2, 0.3 and 0.5) were synthesized by sealed tube method using high purity Pr, $Pr_6O_{11}$, $Co_3O_4$, FeAs, FeP and As. FeAs was obtained by heating Fe and As chips in evacuated sealed silica tubes at 850 °C for 36 hours. FeP was obtained by heating Fe and P powder in evacuated sealed silica tubes at 700 °C for 48 hours. The reactants were weighed in stoichiometric amounts and sealed in evacuated silica ($10^{-4}$ torr) tubes and heated at 950 °C for 24 hours. The resulting powder was ground and compacted into disks. The disks were wrapped in Ta foil and sealed in evacuated silica ampoules and sintered at 1100 °C for 48 hours and then cooled to room temperature. All the chemical manipulations were performed in a nitrogen-filled glove box. The samples were characterized by powder X-ray diffraction

(XRD) using Cu-$K_\alpha$ radiation. The lattice parameters were obtained from a least squares fit to the observed $d$ values using PowderCell software[24]. Energy dispersive X-ray analysis data and average grain size of the sintered samples were obtained with a scanning electron microscope (Carl Zeiss EVO 50WDS electron microscope).

Resistivity measurements in the four probe geometry were carried out in an exchange gas cryostat in 4-300 K temperature range. To measure $H_{C2}$, resistivity as a function of temperature ranging from 5 K up to 20 K were measured under magnetic fields up to 12 T. Diamagnetism of the samples was checked using the magnetisation measurements, performed in Cryogenic Inc. make cryogen based vibrating sample magnetometer operating at 20.4 Hz. The AC susceptibility measurements were done on thin bar shaped samples (5mmx2mmx0.5mm) using the AC susceptibility mode of the VSM operating at 941 Hz and AC field of 0.1mT under different external magnetic field up to 16 T. $H_{irr}$ of thin bar shaped samples was obtained from imaginary component of AC susceptibility ($\chi''$). Using $\chi''$ measurements as a function of applied field the critical current density $J_C$ has been obtained as a function of magnetic field. The variation of resistivity in the superconducting transition region, for various applied magnetic fields was used to obtain the flux pinning potential[25]. The pinning potential was also obtained from the measured variation of the out of phase component of AC susceptibility as a function of frequency of the applied AC field [20].

**III. Results and Discussion**

Figure 1a shows the powder x-ray diffraction patterns of $PrOFe_{0.9}Co_{0.1}As_{1-x}P_x$. Majority of the observed reflections could be satisfactorily indexed on the basis of tetragonal cell (space group *P4/nmm*). Minor amount of $Fe_2As_4O_{12}$ (~5 %) and $FePO_4$ (~5 %) were observed as secondary phases in some compositions. Variation of lattice parameters with phosphorous content(x) in $PrOFe_{0.9}Co_{0.1}As_{1-x}P_x$ is shown in figure 1b. Decrease in both the lattice parameters ( **a** and **c**) was observed which is expected as the ionic size of $P^{3-}$ ion is smaller as compared to $As^{3-}$ ion. Energy-dispersive (EDX) microanalysis of 'x' = 0.1 composition shows the presence of P in addition to Pr, Fe, Co, As and O. Small amount of Si was also seen in the EDX spectra possibly due to the formation of SiO vapour from the quartz tube employed for synthesis or from the Si detector of the microscope. The average grain size for each P substituted sample was evaluated from SEM micrographs obtained under a magnification of ~$10^4$. The grain sizes were determined to be in the 1.8 to 1.4 micron range.

Figure 2a and 2b shows the variation of resistivity and zero field susceptibility with respect to temperature for different P substitution levels. The superconducting onset temperature as measured by resistivity, of the optimal $PrOFe_{0.9}Co_{0.1}As$ sample was ~14.2 K[22]. It is clear from the figure that the normal state resistivity changes to a linear T behaviour with increase in P substitution. The corresponding diamagnetic signals obtained as a function of P substitution are shown in figure 2b. Inset of figure 2b shows $T_C$ versus P fraction obtained from the onset of the diamagnetic signals. The $T_C$ variation obtained from resistivity and susceptibility are similar. $T_C$ increases slightly up to x=0.1 and then decreases at a rate of ~0.13 K/at.% of P substitution. The variation in $T_C$ with P composition seems to be correlated with the variation seen in the a- lattice parameter (Fig.1b). A distinct jump in the a-lattice parameter is seen just above x=0.1, the concentration at which $T_C$ starts decreasing. The observed decrease in $T_C$ with P substitution in optimally doped $PrOFe_{0.9}Co_{0.1}As$[22], is in contrast with the $dT_C/dx>0$ obtained due to chemical pressure by partial Y substitution at the La or Ce site in $ReFeAsO_{1-x}F_x$ compound [26, 27] or the observation of $dT_C/dP > 0$ for an electron doped $LaFeAsO_{1-x}F_x$ [28].

The in-phase ($\chi'$) and out-of-phase ($\chi''$) components of AC susceptibility for phosphorus substituted compounds were measured under applied magnetic fields in the range of 0 to 16 T, at temperatures in the vicinity of the superconducting transitions. The variation of $\chi'$ and $\chi''$ as a function of sample temperature, in the 50% P substituted composition, under the different magnetic fields indicated are shown in figure 3a and 3b. From the figures it can be perceived that field induced broadening of the transitions is smaller in these compounds when compared with earlier such reports on fluorine doped, 1111 compounds [9, 29, 30]. This small change in the superconducting transition width seen with external magnetic field is a feature that was earlier observed in electron doped cuprates [20, 31].

The variation of $(\chi''(T)|_H)$ for 50% P substituted sample, shown in figure 3(b), indicates that the width of the peak remains almost constant but height increases with applied magnetic field. In figure 4 we have plotted the peak temperatures of $\chi''(T)$ as a function of the applied $H_{DC}$, which corresponds to the $H_{irr}$, viz., the field at which magnetic irreversibility sets in. It is seen from figure 4 that there is a systematic increase in $H_{irr}$ with decrease in temperature. In the inset of figure 4 the variation of the irreversibility line $H_{irr}$ with reduced temperature $t=T/T_C$ is shown for different P-substituted compositions. The curves follow a power law $H_{irr}(t) \approx H_{irr}(t=0)*(1-t)^n$ with n varying from 1.7 to 2.3. This value of n is close to the value reported for electron doped single and polycrystalline cuprates, which was n~2 [20]. It is also evident from figure 4 that $H_{irr}(t=0)$ is found to show a large decrease with increase in P substitution.

To evaluate the effect of magnetic field on the resistivity behaviour of the superconducting transitions, the superconducting transitions were measured in fields up to 12 T. The superconducting transitions traced under the application of different external magnetic fields, is shown in figure 5, for some of the P-substituted compounds. It can be seen on careful perusal of each data set that the onset temperature shifts weakly with magnetic field but the zero resistance temperature shifts to a larger extent. The superconducting transition is thus found to broaden with the application of magnetic field. The magnitude of broadening of the transitions gives an indication that the pinning potential in these compounds is rather small [31]. To obtain the variation of the upper critical field ($H_{C2}$) as function of temperature, the $T_C$ at each field was identified as the temperature at which the resistance drops to 90% of the normal state resistivity. The H-T phase diagram for some P-substituted compounds is shown in figure 6. Using the single band WHH formula [32] the zero field upper critical field ($H_{C2}(0)$) was calculated for these P-substituted compounds. The $H_{C2}(0)$ is found to decrease from 64 T for the pristine sample to 29 T for a sample in which 30% of As is substituted by P. For comparison $H_{irr}$-T for the same samples are also shown in the figure 6. It is clear from the figure that at a given temperature the $H_{c2}$ and $H_{irr}$ are widely different for each composition. Further it is apparent from the figure that this difference is not significantly altered by different P substitutions. This result is in stark contrast to observations made on several other 1111 compounds [29, 30] where $H_{irr}$ is very close to $H_{C2}$. The widely different magnitudes of $H_{C2}$ and $H_{irr}$ seen in the present set of samples imply that the flux pinning potential in these compounds is small.

Polycrystalline samples of the present compounds had a ferromagnetic impurity phase, which precluded measurement of $J_C$ by M versus H measurement. $J_C$ was instead obtained from the AC susceptibility measurements carried out as a function of DC magnetic field, for different AC field amplitudes. In the critical state model, the peak in $\chi''(H)$ of the superconducting sample gives an estimate of current density $J_C$ [17, 33,34] according to relation,

$$\mu_0 H_{AC} = \sqrt{2}\mu_0 J_C R \qquad (1)$$

In the above equation, $\mu_0 H_{AC}$ is the amplitude of the applied AC field for which a peak was observed in $\chi''(H)$ and $\mu_0$ is the permeability. Here 'R' is assumed to be the average grain diameter in the case of sintered samples. Further, $J_C(H_{DC}+H_{AC})$ is taken to be independent of AC driving field since $H_{DC} >> H_{AC}$ ($\mu_0 H_{AC} \sim$ 0.1 mT). The AC fields used, ranged from 0.025mT to 0.2mT and were applied along the direction of the DC field. The frequency of the ac field was 941 Hz. Inset of figure 7 shows a plot of normalized $\chi''$ as function of DC bias field for sample with a P fraction of x=0.2, measured at 8 K. From the particular value of AC magnetic field and the grain diameter 'R', $J_C(H)$ was calculated for all the samples using Eq.1, where H is the DC field at which a peak in $\chi''(H)$ plot occurs. From similar measurements done on several P substituted samples, $J_C(H)$ was obtained for different phosphorus substitutions and is shown in figure 7. It can be clearly observed from the figure that Jc (H) decreases precipitously with phosphorus substitution.

The $J_C(H)$ of a superconductor depends on T, H and pinning energy[35]. To identify the origin of the decrease in $J_C(H)$ with P substitution the pinning potential and its variation with field was measured from the magnetic field dependence of resistive transitions to the superconducting state. In the thermally activated flux flow model [36], the variation of the resistivity in the SC transition region can be described by the relationship $R(T,H) = R_0 \exp(-U(T,H)/k_B T)$, where $R_0$ is a constant and U(T,H) is the energy for de-pinning the vortex bundle. U(T,H) is weakly dependent on temperature and is given by $U(T,H)=U_0(H)(1-t)$, where $t=T/T_C$. Therefore the effective pinning energy U(T, H) is expected to be same as U(T=0, H) [31]. To obtain the pinning potential from the experimental data (R(T,H)) the slope of ln $R$ versus 1/T was obtained for each DC field, H for each P-substituted compound. From magneto-resistance data, U(H) was evaluated for three P fractions. The variation of U as a function of applied field obtained for these P concentrations is displayed in figure 8. The data can be described by a power law of the form $U(H) \approx U_0 H^n$ with $n \approx 0.5$ to 0.8. The value of the n is similar to that reported for electron doped cuprates[20, 30] but is very high compared to n=0.09 for K doped $BaFe_2As_2$ single crystals[25]. The pinning potential of the present compounds reduces to very low values with increase in magnetic field. The magnitude of $U_0 \sim$ 45 meV obtained from fits (cf. figure 8) for these compounds is very low as compared to 774 meV obtained for potassium doped Ba122[24] or 172 meV obtained for F doped Nd1111 system [37]. It is clear from the plot of U(H) that substitution of P does not significantly alter the flux pinning potential of $PrOFe_{0.9}Co_{0.1}As$ and thus cannot be the reason for the drastic reduction of in-field $J_C$. It would be interesting to find out if the small U arises due to Co substitution, by performing similar studies in Co doped samples viz., $PrOFe_{1-x}Co_xAs$ for various x.

To verify on the reliability of pinning potentials obtained from R(T) data, the frequency dependence of the peak position of $\chi''(T)$ was also studied for the x=0.1 composition, under fixed magnitude of external DC and AC magnetic fields. $\chi''(T)$ was measured from 93 Hz to 9093 Hz at an external magnetic field of 1 T and a fixed magnitude of a parallel AC field of 0.1mT. Inset of the figure 8 shows a plot of $\chi''$ (normalized to its maximum value) as a function of temperature, measured at different frequencies. It is clearly seen from inset of figure 8, that the peak of $\chi''$ shifts towards higher temperature with increase in measuring frequency. Similar sets of measurements were performed for several external DC magnetic fields up to 11 T. The pinning potential (U) at a given field was calculated using the slope of $\ln(\omega)$ versus $1/T_p$, where $\omega$ is the frequency of the AC field employed and $T_p$ the peak temperature. U(H) variation obtained from ac susceptibility measurements is shown in figure 8 along with the results obtained from resistivity measurements. It is evident from the figure that the values of U (H) and n found from measurements of $\chi''$ are consistent to the values obtained from Arrhenius fits of R(T) data.

## IV. Conclusions

The substitution of phosphorus at the As site of $PrOFe_{0.9}Co_{0.1}As$ superconductor reduces the transition temperature of the optimally doped compound at a rate of ~0.13K/at.%. Whereas, the maximum reported decrease rate of ~ 9 K/at%, is seen for the case of Zn substitution at Fe site in $LaFeAsO_{0.85}$[38]. The P substitution also changes the normal state resistivity behaviour of the parent compound and the normal state resistivity becomes nearly linear with temperature. The $H_{C2}$ is also found to decrease with the P substitution and reduces to a value of 29 T for the x=0.3 composition as compared to 63 T seen in the x=0.0 sample.. The large difference between $H_{irr}$ and $H_{C2}$ points to very small pinning potential, which is confirmed by the values obtained by Arrhenius fits of in-field resistive transitions and frequency dependence of $\chi''(T)_{max}$. The pinning potential is found to be a factor of four times smaller than that reported for other F-doped oxypnictide compounds. Further the pinning energy, does not vary significantly with P substitution indicating that it has little or no contribution in the drastic reduction of the in-field critical current density $J_C(H)$.

The observed decrease in $T_C$ with P substitution, despite the decrease in lattice volume, is in distinct contrast to the expectation in the oxyarsenides, especially in the light of the positive dTc/dP of ~5 K/Gpa seen in $LaOFeAs_{0.9}F_{0.1}$[28]. The most comprehensive study of the effect of pressure on the Tc has been carried out for the $BaFe_{2-x}Co_xAs_2$ system, where a positive $dT_C/dP$ has been observed for all Co concentrations[40]. Although P substitution leads to a reduction in lattice volume and can be treated equivalent to the application of pressure; local variations in the Fe-pnictogen bonds lengths are inevitable and can given rise to disorder induced scattering potentials, although expected to be weak[41] in the case of P substitution. Recent calculations[42] for the related oxyarsenide, LaOFeAs, indicate that the pnictogen height with respect to the Fe layer can swap the ground state from a high $T_C$ fully gapped state to a low $T_C$ nodal pairing state. From the correlated $T_C$ and a&c lattice parameter variations seen in our data (see Fig1b and insets in Fig.2) it is evident that $T_C$ depletes beyond ~10 % P substitution, a concentration at which the a- lattice parameter also shows a marked change. Since the a &c lattice constants would determine the pnictogen position in the structure, the electronic structure could be altered at this composition, consequently resulting in a reduced superconducting pairing and $T_C$. Since $H_{C2}$ and $J_C$ depend on $T_C$, these superconducting properties could also show a deterioration with P substitution. Poor flux pinning due to the polycrystalline nature of the samples can also contribute to the low $J_C$ and $H_{irr}$ observed in these samples.

**Acknowledgements:**
The authors thank Dr. Shamima Hussain and Dr. G. Amarendra of UGC-DAE CSR, Kalpakkam node for the timely grain size measurements. AB thanks Dr. Y. Hariharan, a former colleague, for a critical reading of the manuscript. AKG thanks DST, Govt. of India for financial support. JP and GST thank CSIR, Govt. of India, and DST for fellowships, respectively.

**References:**

1. Kamihara Y, Watanabe T, Hirano M and Hosono H 2008 *J. Am. Chem. Soc.* **130** 3296

2. Chu C W 2009 *Nature Physics* **5** 787-789

3. Cvetkovic Vand Tesanovic Z 2009 *Europhysics Letters* **85** 37002

4. Li T 2008 arXiv:0804.0536v1 [cond-mat.supr-con]

5. Kuroki K, Onari S, Arita R, Usui H, Tanaka Y, Kontani H and Aoki H 2008 *Phys. Rev. Lett.* **101** 087004

6. Ma F and Lu Z Y 2008 *Phys. Rev. B* **78** 033111

7. Yamamoto A et al 2008 *Appl. Phys. Lett.* **92** 252501

8. Wang L, Gao Z, Qi Y, Zhang X, Wang D and Ma Y 2009 *Supercond. Sci. Technol.* **22** 015019

9. Hunte F, Jaroszynski J, Gurevich A, Larbalestier D C, Jin R, Sefat A S, McGuire M A, Sales B C, Christen D K and Mandrus D 2008 *Nature* **453** 903

10. Wang Z-S, Luo H Q, Ren C, and Wen H H 2008 *Phys. Rev. B* **78** 140501

11. Altarawneh M M, Collar K and Mielke C H 2008 *Phys. Rev. B* **78** 220505

12. Chong S V, Hashimoto S and Kadowaki K 2010 *Solid State comm.* **150** 1178

13. Qi Y, Wang L, Wang D, Zhang Z, Gao Z, Zhang X and Ma Y 2010 *Supercond. Sci. Technol.* **23** 055009

14. Gao Z , WangL , Qi Y , Wang D , Zhang X and Ma Y 2008 *Supercond. Sci. Technol.* **21** 105024

15. van der Beek C J and Kes P H 1991 *Phys. Rev. B* **43** 13032

16. Civale L et al. 1991 in *Magnetic Susceptibility of Superconductors and other Spin Systems* ed Hein R A (Plenum New York) p 313

17. Gomory F 1997 *Supercond. Sci. Technol.* **10** 523

18. Malozemoff A P, Worthington T K, Yeshurun Yand Holtzberg F 1988 *Phys. Rev. B* **38** 7203

19. van den Berg J, van der Beek C J, Kes P H, Mydosh JA, Menken M J V and Menovsky A A 1989 *Super. Sci. Technol.* **1** 249

20. Fabrega L, Fontcuberta J, Civale L and Pinol S 1994 *Phys. Rev. B* **50** 1199

21. Kasahara S, Shibauchi T, Hashimoto K, Ikada K, Tonegawa S, Okazaki R, Shishido H, Ikeda H, Takeya H, Hirata K, Terashima T and Matsuda Y 2010 *Phys. Rev. B* **81** 184519

22. Prakash J, Singh S J, Das D, Patnaik S, Ganguli AK, 2010 *Journal of Solid State Chemistry* **183** 338

23. Lorenz B, Sasmal K, Chaudhury R P, Chen X H, Liu R H, Wu T and Chu C W 2008 *Phys. Rev. B* **78** 012505

24. Kraus W, Nolze G *PowerCell for Windows version 2.4* 2000 Berlin Germany

25. Wang X L et al 2010 *Phys. Rev. B* **82** 024525

26. Prakash J, Singh S J, Patnaik S and Ganguli A K 2010 *J. Solid State Chem.* **183** 338

27. Tropeano M, Fanciulli C, Canepa F, Cimberle M R, Ferdeghini C, Lamura G, Martinelli A, Putti M, Vignolo M and Palenzona A 2009 *Phys. Rev. B* **79** 174523

28. Takahashi H, Igawa K, Arii K, Kamihara Y, Hirano M and Hosono H 2008 *Nature* **453** 376

29. Jaroszynski J, Riggs S C, Hunte F, Gurevich A, Larbalestier D C and Boebinger G S 2008 *Phys. Rev. B* **78** 064511

30. Moll P J W, Puzniak R, Balakirev F, Rogacki K, Karpinski J, Zhigadlo N D and Batlogg B 2010 *Nat. Mat.* **10** 628

31. Fabrega L, Fontcuberta J and Pinol S 1993 *Phys. Rev. B* **47** 15250

33. Werthamer N R, Helfand E and Hohenberg P C 1966 *Phys. Rev.* **147** 295

34. Chen D X, Nogues J and Rao K V 1989 *Cryogenics* **29** 800

35. Gomory F 1989 *Solid State Commun.* **70** 879

36. Wang X L, Dou S X, Hossain M S A, Cheng Z X, Liao X Z, Ghorbani S R, Yao Q W, Kim J H and Silver T 2010 *Phys Rev B* **81** 224514

37. Kes P H, Aarts J, van den Berg J, van der Beek C J and Mydosh J A 1989 *Super. Sci. Technol.* **1** 242

38. Jaroszynski J, Hunte F, Balicas L, Jo Y-jung, Raicevic I, Gurevich A and Larbalestier D C 2008 *Phys. Rev. B* **78** 174523

39. Guo Y F et al 2010 *Phys. Rev. B* **82** 054506.

40. Ahilan K, Ning F L, Imai T, Sefat A S, McGuire M A, Sales B C, and Mandrus D *Phys Rev B* 2009 **79**, 214520

41. Lina E. Klintberg, Swee K. Goh,, Shigeru Kasahara, Yusuke Nakai, Kenji Ishida, *Journal Phys. Soc Japan* 2010 **79** 123709

42. Kuroki K, Usui H, Onari S, Arita R and Aoki H, 2009 *Phys. Rev B* **79** 224511

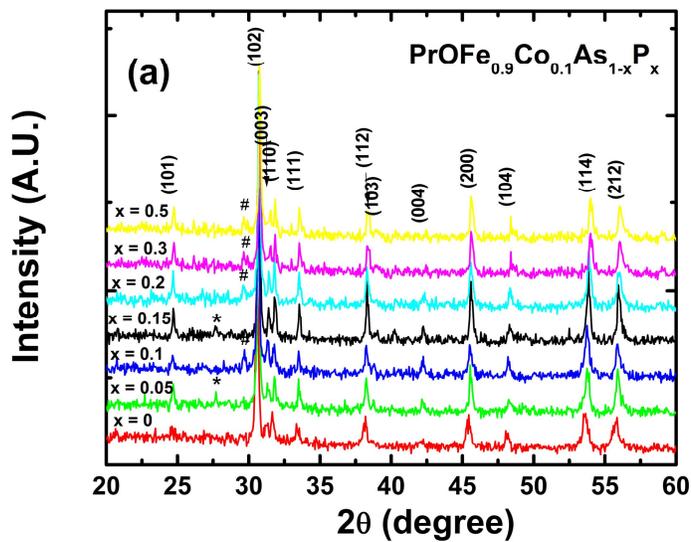

**Figure 1a** Powder X-ray diffraction patterns (PXRD) of $PrOFe_{0.9}Co_{0.1}As_{1-x}P_x$ sintered at 1100 ºC. The impurity phases are marked by the symbols shown in parenthesis, $Fe_2As_4O_{12}$(*), $FePO_4$ (#).

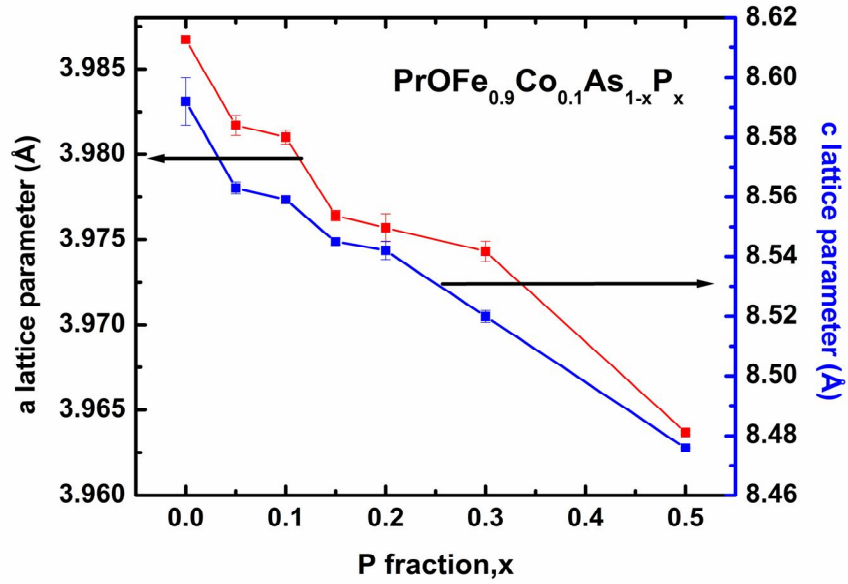

**Figure 1b** Variation of lattice parameters with P content in PrOFe$_{0.9}$Co$_{0.1}$As$_{1-x}$P$_x$. The lattice parameters for x=0.0 are taken from ref. 22.

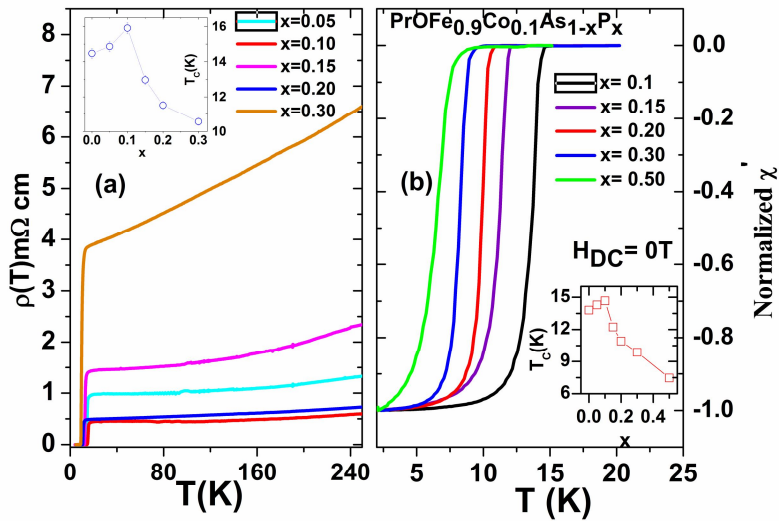

**Figure 2a**: Resistivity versus temperature for different P fractions in PrOFe$_{0.9}$Co$_{0.1}$As$_{1-x}$P$_x$, Inset: Decrease in T$_C$ due to P substitution from ρ(T). **Figure 2b**: χ′(T) normalised to the value at the lowest temperature for each P fraction. Inset: Decrease in T$_C$ due to P substitution from onset of diamagnetism obtained from χ(T). T$_C$ values for x=0.0 in the insets are taken from ref.[22].

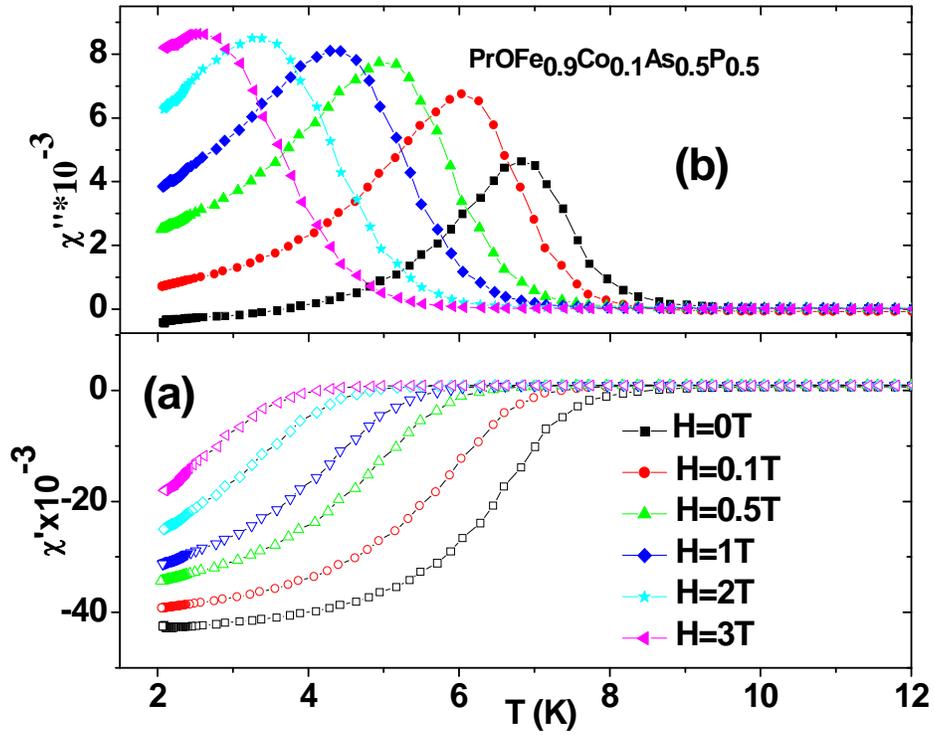

**Figure 3:** Variation of $\chi'$ and $\chi''$ with temperature for different DC magnetic fields indicated for the x=0.5 composition.

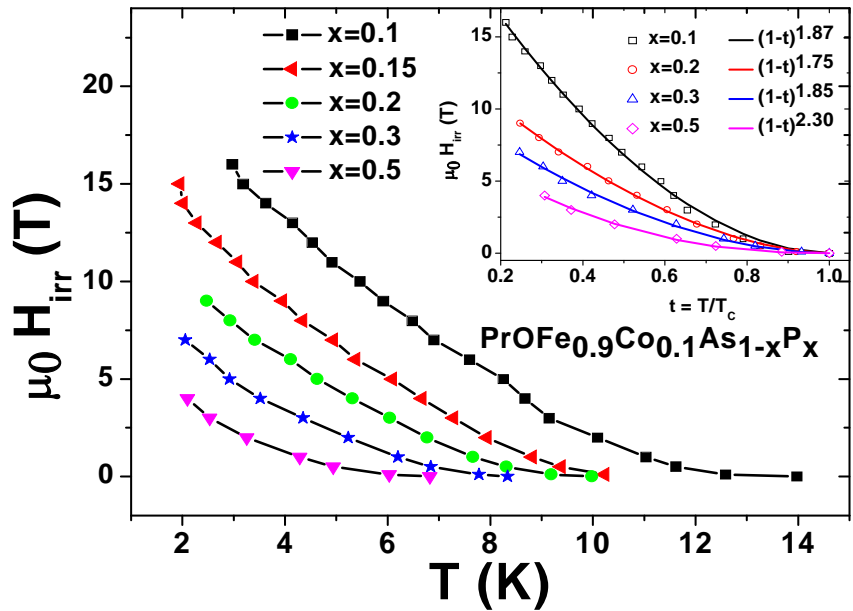

**Figure 4**: $H_{irr}$-T phase diagram for different P concentrations. Inset: Irreversibility line and its fit to $(1-t)^n$ law.

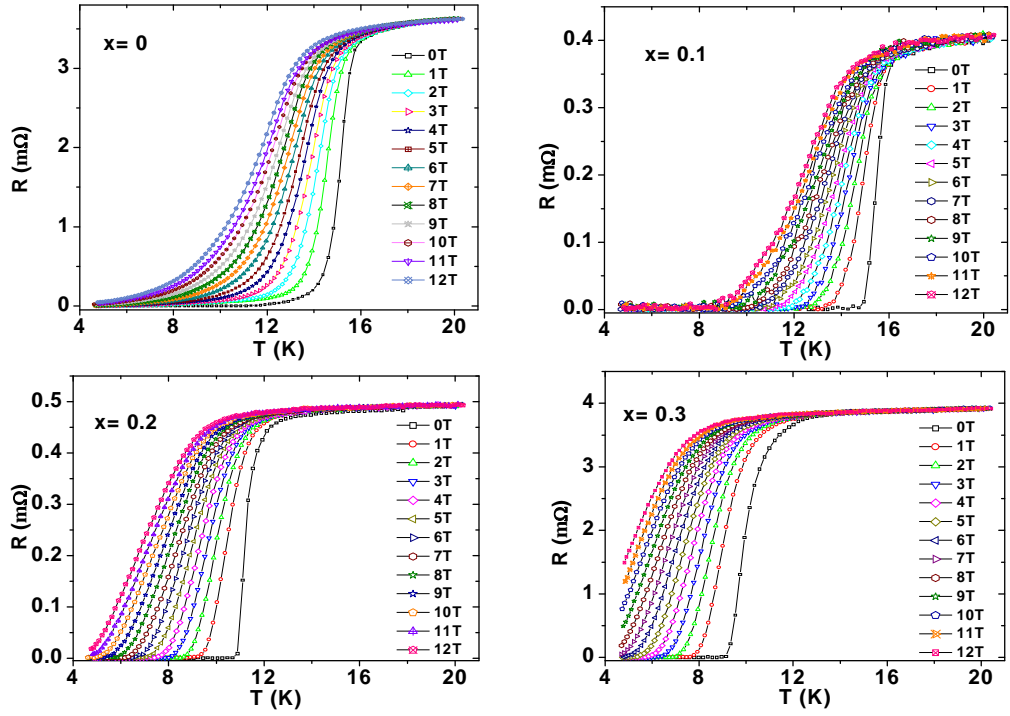

**Figure 5**: Magnetoresistance data for PrOFe$_{0.9}$Co$_{0.1}$As$_{1-x}$P$_x$ for x=0, 0.1, 0.2, 0.3 compositions.

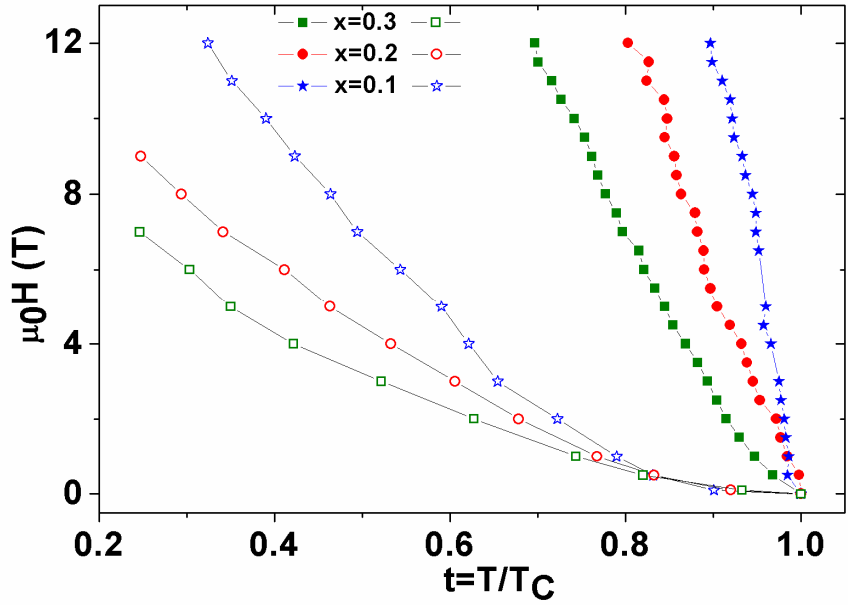

**Figure 6**: Upper critical field H$_{C2}$(solid) and H$_{irr}$(open) as a function of reduced temperature.

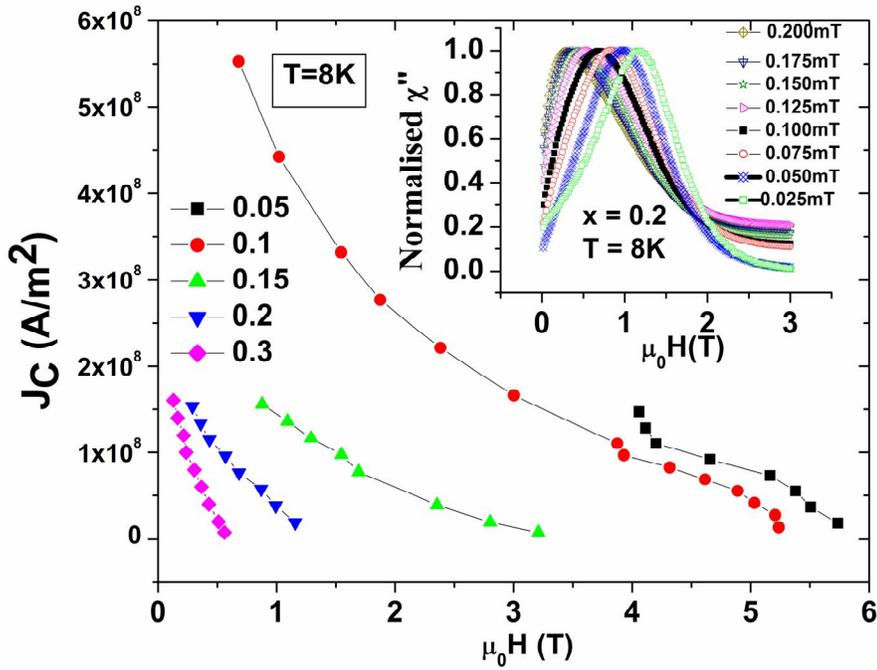

**Figure 7**: In-field intra-grain current density $J_C(H)$ measured at 8 K for different P fractions in PrOFe$_{0.9}$Co$_{0.1}$As$_{1-x}$P$_x$; inset: Variation of $\chi''$ with DC magnetic field at different AC fields applied parallel to DC field, measured at 8 K, for a P fraction x=0.2.

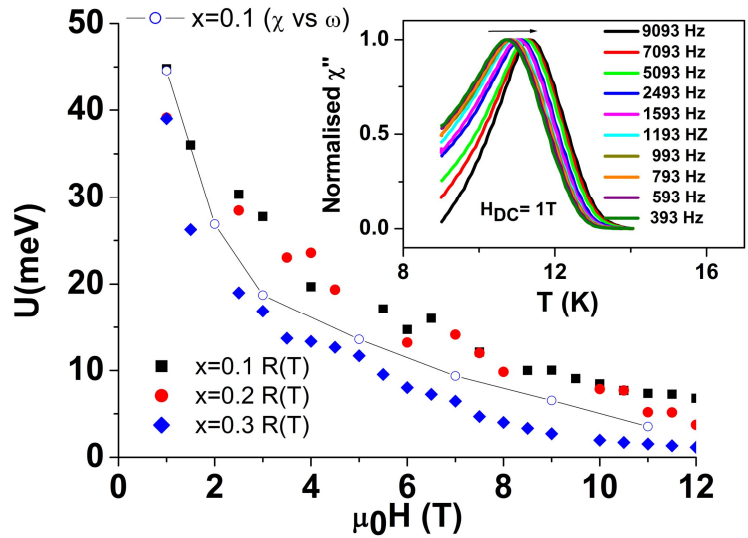

**Figure 8**: Pinning potential variation with magnetic field as calculated from R(T) measurements at different magnetic fields for different P fractions in PrFe$_{0.9}$Co$_{0.1}$As$_{1-x}$P$_x$. Open circles are values obtained from $\chi''(T)$ measured for various frequencies of AC field. Inset shows $\chi''(T)$ for the sample with x=0.1 measured for different frequencies of the AC field under a DC field of 1T, the arrow indicates increasing frequency.